\documentclass[twocolumn,aps,prl]{revtex4}
\usepackage{graphicx}
\usepackage{amssymb}

\begin{document}
\title{Dynamics of Surfactant Driven Fracture of Particle Rafts}
\author{Dominic Vella$^{1,2}$}
\author{Ho-Young Kim$^{1,3}$}
\author{Pascale Aussillous$^2$}
\author{L. Mahadevan$^1$}
\email{lm@deas.harvard.edu}
\affiliation{$^1$ Division of Engineering and Applied Sciences, Harvard University, Pierce Hall, 29 Oxford Street, Cambridge MA 02138}
\affiliation{$^2$ITG, Department of Applied Mathematics and Theoretical Physics, University of Cambridge, Wilberforce Road, Cambridge, CB3 0WA,  U.K.}
\affiliation{$^3$School of Mechanical and Aerospace Engineering, Seoul National University,
Seoul 151-744, Korea}

\begin{abstract}
We investigate the dynamic fracture of a close-packed monolayer of particles, or particle raft, floating at a liquid-gas interface induced by the localised addition of surfactant. Unusually for a two-dimensional solid, our experiments show that the speed of crack propagation here is not affected by the elastic properties of the raft. Instead it is controlled by the rate at which surfactant is advected to the crack tip by means of the induced Marangoni flows. Further, the velocity of propagation is not constant in time and the length of the crack scales as $t^{3/4}$. More broadly, this surfactant induced rupture of interfacial rafts suggests ways to manipulate them for applications.
\end{abstract}

\date{February 7, 2006}
\maketitle

The curious behavior of particulate interfaces that separate liquids is a source of interesting questions at the intersection of hydrodynamics and elasticity. For instance, liquid drops coated with a fine hydrophobic powder become non-wetting \cite{Aussillous}, forming an artificial analog of a much older solution stumbled upon by insects \cite{Pike}. Similarly, the addition of particles to the surface of liquid drops prior to coalescence stabilises the coalesced drops to the common pinch-off instability  and can lead to reversible morphological instabilities such as buckling when subject to pressure \cite{Stancik}. Particle covered interfaces also occur at an intermediate stage during the production of {colloidosomes} \cite{Dinsmore}, armored bubbles \cite{Anand} and porous particle aerosols for drug delivery \cite{Edwards}. From both a fundamental and technological point of view these particulate interfaces pose a number of problems and opportunities. Understanding their rheological properties has the potential to open up ways of controlling them, but remains mostly an open question. 

\begin{figure}
\centering
\vspace{0.3cm}
\includegraphics[height=10cm]{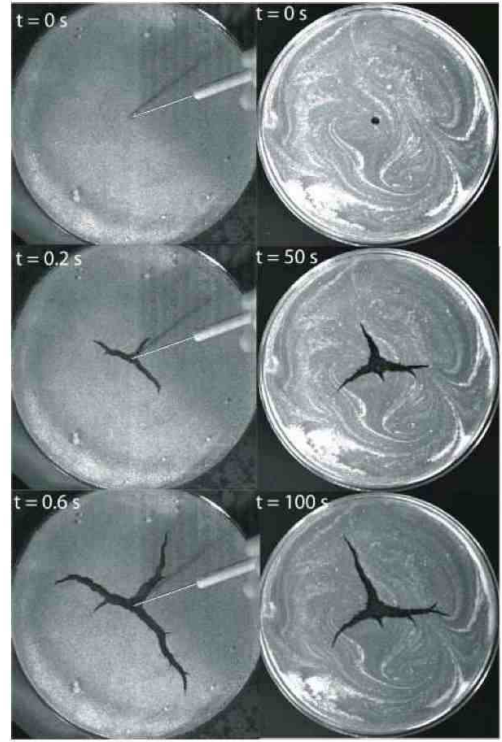}
\caption{Time series showing the branching fracture of an interfacial particle raft consisting of $100 \mu m$ Pliolite particles floating at an air--water (left) and air--glycerol (right) interface. The diameter of the circular dish in each case is $15$ cm, and the underlying fluid layer is $2$ cm thick.}
\label{tseries1}
\end{figure}

Recent experiments \cite{Cicuta,Vella} show that a densely packed
monolayer of particles at an interface (a `particle raft') has many
of the characteristics of a two-dimensional linear elastic solid in
certain regimes. For example, under compressive loading, a particle
raft statically buckles out of the plane demonstrating that
collectively its constituent particles possess a non-zero shear
modulus $G \sim  \gamma/d$ \cite{Vella}, with $\gamma$
  the surface tension coefficient of the pure liquid-gas interface
and $d$  the particle diameter. This ability to sustain finite shear
stresses arises from a combination of capillary forces and the short
range steric constraints due to particle-particle contact, and is seen
in a variety of similar systems such as armored bubbles, drying
colloidal drops and suspensions \cite{Anand,Tsapis, Pauchard,
  Dufresne}, although understanding the kinetics of onset of this
elastic behavior constitutes work in progress.  Going beyond the
linear elastic behavior,  the ability to sustain finite shear stresses
suggests that these rafts should also be able to sustain fractures
which relieve stresses primarily in one direction \cite{Vella}. This
fracture can be observed in the particulate scum that forms on the surface of a
cup of black tea, which is then fractured by the addition of milk. A similar phenomenon is observed in pond scum when fractured by the ripples induced by a pebble. Here we use an interfacial particle raft as a vehicle to study the cracks induced by the addition of surfactant  and provide a model for their unusual dynamics. This study provides a window on the
nonlinear rheology of these fluid-solid interfacial composites. As we
will see, the dynamics of crack growth is governed by the rate at which
surfactant can be supplied to the tip. 


Our experimental system is a densely packed monolayer of non-Brownian
particles  \cite{Vella,pliolite} trapped at the interface between air and a water-glycerol mixture contained within a circular dish of diameter $15$ cm and depth $2$
cm. A drop of surfactant (whose volume does not affect the results) is introduced somewhere in the layer with a needle, which when clean is benign, i.e.~it does not open a crack. We used Polyoxyethylene Sorbitan Monooleate (EM Science) as the surfactant, though household detergents have the same effect. The localized reduction of surface tension causes a tensile stress in the
particulate layer; a crack nucleates at the needle, grows and
eventually branches, as shown in fig.\ \ref{tseries1}. The composition of the water--glycerol mixture was varied allowing us to vary the viscosity of the underlying liquid, $\mu$, from $10^{-3}-0.3\mathrm{~Pas}$. This variation has a pronounced effect on the speed of propagation of the crack (note the disparate timescales in fig.~\ref{tseries1}) so that the typical propagation speeds in our system lie in the range $10^{-3}-10^{-1}\mathrm{~ms^{-1}}$. These cracks thus travel at speeds of between $0.2\%$ and $20\%$ of the shear wave speed, which scales as   
\begin{equation}
v_s \sim \sqrt{\frac{G}{\rho}}\sim\sqrt{\frac{\gamma}{2\rho(1+\nu) d}}\sim 0.5 \mathrm{ms}^{-1},
\end{equation} where we have used the estimate for the shear modulus $G$ and Poisson ratio $\nu$ determined and verified in \cite{Vella}, and $\rho$ was taken as the density of water and the (neutrally buoyant) Pliolite. Over the large range of crack speeds in our experiment, we  observe phenomena  such as crack branching, kinking and the appearance of frustrated side--branches, which makes our system rather unusual. Indeed, they question the hypothesis that dynamical effects alone can give rise to these phenomena and suggest that instead heterogeneities may be the dominant cause of these effects. As an example, we consider the process that leads to the kinking of the crack path which seems to be governed by a simple phenomenological rule: the crack propagates in a particular direction through a grain of the solid until it reaches the grain boundary. At this point it changes direction to travel in whichever of the two possible cleavage directions is closest to its current direction of travel. Understanding this requires a consideration of the effects of disorder; indeed our experiments show quite clearly that crack branching and kinking are not due to inertial effects alone, contrary to current thinking on the subject. 

\begin{figure}
\centering
\vspace{0.3cm}
\includegraphics[height=6.5cm]{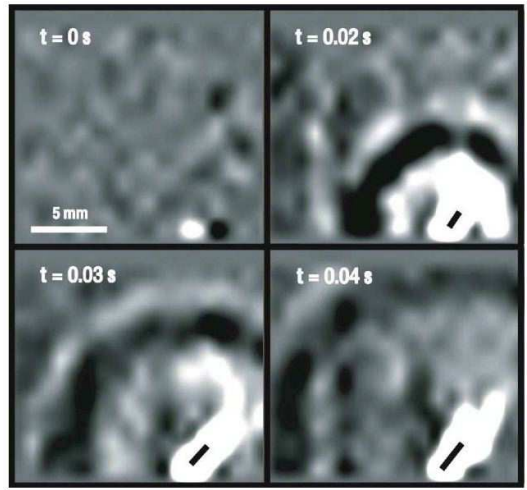}
\caption{Time series of the two--dimensional divergence of the velocity field in a raft of $90\mu$m particles showing the propagation of a sound wave (circular wave fronts) shortly after the injection of surfactant. The speed of this wave agrees well with that based on the estimate of the Young's modulus obtained by buckling experiments \cite{Vella}. The orientation and length of the crack is represented by the solid black lines in each of the last three frames.}
\label{s_wave}
\end{figure}

The fracture of these particle rafts was observed using a high speed CCD camera at frame rates of up to $200\mathrm{s}^{-1}$. The images produced were subsequently analysed using image analysis software (DigiFlow, DL Research Partners) to determine the velocity field in the raft and the divergence of this velocity field. Typically the time series of the divergence field in the raft at short times (see fig.\ \ref{s_wave} for an example) shows a wave advancing ahead of the crack. The typical speed of this wave is $\sim{\cal O}(0.5)\mathrm{ms}^{-1}$, which is  very close to the speed of shear waves in the solid $v_s$, but which is considerably faster than the speed of the crack itself. Thus inertial effects lead to rapid dynamic equilibrium in the raft which learns of the elastic disturbance due to the presence of the crack acoustically, even though the crack itself propagates much more slowly. 

\begin{figure}
\centering
\vspace{0.3cm}
\includegraphics[height=4cm]{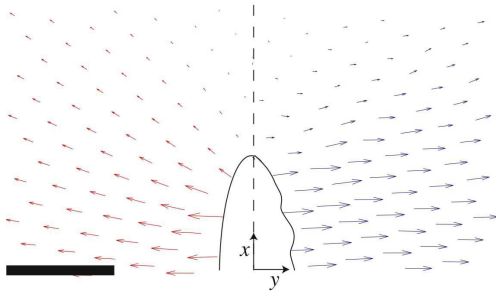}
\caption{The computed (left) and measured (right) velocity fields
  around the crack tip (see text). The arrows illustrate the
  instantaneous velocity at a particular point, but have been scaled
  for clarity and represent the distance that the point would move in the next $0.065\mathrm{~s}$. In the model (see eq.~(\ref{timodisp})), the geometry
  of the crack (solid black line on right) is represented by an elliptical hole of similar
  dimensions. The scale bar represents $5\mathrm{~mm}$.}
\label{timo}
\end{figure}

Image analysis also reveals the spatial inhomogeneity of the response
of the raft. In particular, ahead of an advancing crack there is a
shadow zone where the velocity of the raft is negligible, as shown in
fig.\ \ref{timo}.  Since  the speed of crack propagation is much slower than the speed at which the details of the crack are communicated to the raft and its boundaries, we assume that the displacement within the raft is determined instantaneously by the geometry of the crack. Thus we may turn to the classical plane stress solution for the equilibrium of a plate with  an elliptical hole subject to internal pressure~\cite{Timoshenko, Mushkhelishvili}. We summarize briefly here the approach followed in section 62 of \cite{Timoshenko}. Since the raft is loaded in plane stress, the Airy stress function, $\phi$, satisfies $\nabla^4\phi=0$. Letting $z=x+iy$ where $(x,y)$ is a Cartesian co-ordinate system centred on the hole, we introduce the complex potentials $\psi(z)$ and $\chi(z)$ with $\phi=\mathrm{Re}\bigl(\bar{z}\psi(z)+\chi(z)\bigr)$. The boundary condition that the stress at infinity is zero leads to $\mathrm{Re}(\psi'(z))=\bar{z}\psi''(z)+\chi''(z)=0$ there. At the elliptical boundary of the hole, the stress must equal the spreading pressure, $S$. This condition is simplified by moving into an elliptical co-ordinate system $z=c\cosh\zeta$ with $\zeta\equiv \xi+i\eta$ where $\xi$ is constant on a given ellipse and on the boundary of the hole $\xi=\xi_0$. This leads to the potentials
\begin{equation}
\psi(z)=\frac{S}{2}(c\sinh\zeta-z),\hspace{0.5cm}\chi(z)=-\frac{S}{2}c^2\zeta\cosh2\xi_0,
\end{equation} which in turn give the displacement field, $(u,v)$, as
\begin{eqnarray}
u+iv&=&\frac{S(1+\nu)}{2E}\left(\frac{3-\nu}{1+\nu}(c\sinh\zeta-z)+z\right.
\nonumber\\
&&-\left.c\cosh \zeta {\coth \overline\zeta}+\frac{c\cosh2\xi_0}{{\sinh\overline \zeta}}\right),
\label{timodisp}
\end{eqnarray} where $\nu$ is the Poisson ratio ($\nu=1/\sqrt{3}$ for
hexagonally close-packed spheres interacting via a central
force\cite{Vella}). We may then obtain a theoretical estimate of the
velocity field by using the measured crack shapes at two
instants of time to determine the displacement fields from
(\ref{timodisp}) and subtracting them. The results of this are shown
in fig.\ \ref{timo}, where a value of $S\approx 0.3\gamma/d$ has been used to fit theory to experiment. Integrating this fitted value of $S$ over the thickness of the layer shows that the difference in surface tensions between the crack and the solid is $0.3\gamma$. This is indistinguishable from the directly measured difference in interfacial tensions between pure and contaminated interfaces. The agreement between computation and experiment shown in fig.\ \ref{timo} is reasonable given the simplicity of our model and the departure of the real crack from our elliptical idealisation. Over the points shown in fig.~\ref{timo}, the average error is $30\%$, while the angle of the shadow zone ahead of the crack tip is similar in both instances.

We now move towards a quantitative description of the dynamics of a
single crack by measuring its length, $L_c$, as a function of time,
$t$, following its initiation. It is observed experimentally that the cracks terminate at some final length, $L_\infty$, which does not vary between experiments in a systematic manner. 
This suggests that the heterogeneity of packing, the initial configuration of the raft and the proximity of other cracks may influence the propagation of a single crack.  We note that after compressing the raft to the point at which
buckling sets in and then adding surfactant, no crack was formed:
 if there is insufficient free area within the raft for
particle rearrangements to occur, the crack will not propagate at
all. Conversely, once a crack has reached its equilibrium length, any
further compression of the raft (in the direction perpendicular to the
crack) leads immediately to the onset of buckling. In addition, the
crack maintains its final shape for hours after the crack has ceased
to propagate. This suggests that, as the crack propagates, the
particles are consolidated to liberate liquid-gas surface area at the
crack, while after the crack has progressed, the particles within the
raft are in a stable jammed state compared to that prior to crack initiation.

\begin{figure}
\centering
\includegraphics[height=5.2cm]{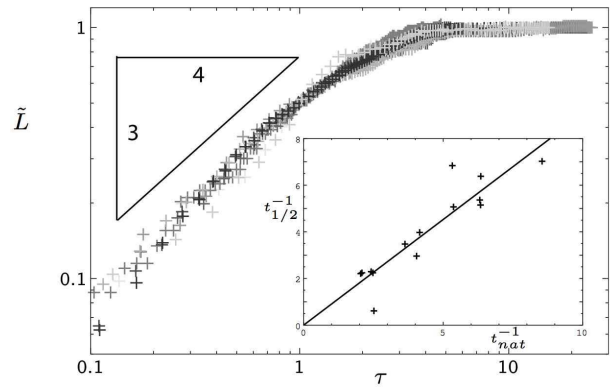}
\caption{Main figure: Non-dimensional crack length as a function of non-dimensional time for various viscosities and particle sizes (points). Inset: Experimental confirmation of  the scaling $t_{1/2}^{-1}\sim t_{nat}^{-1}$ giving the growth rate in terms of the different experimental parameters as in (4).}
\label{clength}
\end{figure}

The separation of speeds between the propagation of the crack and the  sound waves shown in fig.\ \ref{s_wave} suggests that the fracture itself is limited by the rate of advection of surfactant to the crack tip. To test this possibility, we used neutrally buoyant food coloring as the surfactant and observed that throughout the motion, the dye was confined to the crack area and that cracks only propagated with visible amounts of dye at the tip. This suggests that the spreading of surfactant is indeed vitally important for crack propagation. We now move on to consider the dynamics of this process.

As the crack propagates, energy is liberated by the reduction in
surface tension of the crack area compared to the pure liquid. The
dynamics of the crack propagation are thus  governed by the balance
between the rate at which this energy is produced and the rate at
which energy is dissipated by the motion of the raft. There are two
possible dissipation mechanisms. The first of these is the dissipation
that occurs in a Blasius boundary layer, which is caused by the impulsive surface flow induced by  differences in surfactant concentration via Marangoni stresses. A time $t$ after initiation of the crack, the typical vertical extent of this surface boundary layer is $\delta \sim (\mu t/\rho)^{1/2}$ with $\mu$ the dynamic viscosity of the liquid. The rate of viscous dissipation within the volume, $V$, of this boundary layer is ${\cal D}_{bbl}\sim2\mu\int_V(\nabla u)^2dV\sim \mu (\dot{L_c}/\delta)^2L_c^2\delta$. There is another source of dissipation due predominantly to the lubrication flow in the spaces between particles: a typical shear rate, $\dot{\gamma}$, gives a dissipation rate of ${\cal D}_{lub}\sim\mu L_c^2d\dot{\gamma}^2$, an overestimate since it neglects the crucial fact that the liquid between particles can move out of the plane. Assuming that this shear rate is of the same magnitude as the rate of compression of the raft, we find experimentally that $\dot{\gamma}\sim1\mathrm{s}^{-1}$ so that ${\cal D}_{lub}/{\cal D}_{bbl}\lesssim 10^{-5}\ll1$,  i.e. we can neglect the dissipation caused by inter-particle motions and focus on the balance between the liberation of surface energy and dissipation in the sub--surface boundary layer, considered by Hoult and others\cite{Hoult}. Then assuming that the crack area is $A_c$, the rate of energy liberation by the propagation of surfactant scales as $\Delta\gamma \dot{A}_c$, where $\Delta\gamma$ is the change in surface tension coefficient between pure and contaminated interfaces. Balancing this with the rate of energy dissipation in the boundary layer ${\cal D}_{bbl}\sim \mu (\dot{L_c}/\sqrt{\mu t/ \rho})^2A_c\sqrt{\mu t/ \rho }$,  yields
\begin{equation}
L_c\sim\left(\frac{\Delta\gamma^2}{\mu\rho}\right)^{1/4}t^{3/4}.
\label{threefour}
\end{equation} In fig.~\ref{clength}, we show that the scaled crack length $\tilde{L}\equiv L_c/L_\infty \sim \tau^{3/4}$  over more than a decade in the scaled time $\tau =t/t_{1/2}$ where $t_{1/2}$  is the time taken for the crack to grow to half of its final length. Eventually, the crack slows down as it propagates into regions of densely packed particles and finally stops. In the inset to fig.~\ref{clength}, we also see that  $t_{1/2}^{-1}\sim t_{nat}^{-1}$, where $t_{nat} \equiv\left(\mu\rho L_\infty^4/\Delta\gamma^2\right)^{1/3}$ is the natural time-scale from (\ref{threefour}). We observe that the boundary layer extends through the depth, $h$, of the fluid sublayer when $t\gtrsim h^2 \rho/\mu $. For the raft shown in the right-hand panel of fig. \ref{tseries1}, $\mu=1.4\mathrm{~Pas}$ so that this time is ${\cal O}(1\mathrm{~s})$, explaining why this crack propagates even slower than we would expect on the basis of (\ref{threefour}).

Our experiments on the surfactant-induced failure in an interfacial particulate raft show  some unusual features such as crack branching and kinking even though the fractures are limited by the drag induced by the formation of a viscous boundary layer in the underlying fluid, rather than solid inertia. They also confirm the solid-like response of densely packed
particles trapped at a liquid-gas interface. Understanding the means by which fractures occur and the dynamics of their propagation in these systems may have very real
benefits in, for example, stabilising foams, drops and bubbles by the
addition of particles. One particularly interesting possibility lies
in the delivery of drugs by inhalation, where trojan parcels of drugs
in drops are coated with particulate rafts~\cite{Edwards}. Our work
suggests that the delivery process might be expedited by using the surfactant lining of the lungs to induce the rupture of the particle shell. On a different note, this transport-induced weakening and growth of a crack is similar to what occurs in corrosion cracking and hydrogen embrittlement in solids, and thus our system may also provide a
two-dimensional model to study some of these complex phenomena.

\begin{acknowledgments}
Acknowledgments: We thank K. and A. Mahadevan for an entertaining time at the kitchen tables in the two Cambridges that led directly to this series of investigations, and J. Hutchinson for pointing out the possible connection of our model system to corrosion cracking.
\end{acknowledgments}

 \end{document}